\documentclass{aa}
\usepackage{graphics,psfig}
 
\begin{document}      

\title{A search for cold dust around neutron stars}

\author{
        O.~L\"ohmer \inst{1},
        A.~Wolszczan\inst{2,3,1},
        R.~Wielebinski\inst{1}
        }

\institute{Max-Planck-Institut f\"ur Radioastronomie, Auf dem H\"ugel 69,
              D-53121 Bonn, Germany
    \and
           Department of Astronomy and Astrophysics, Pennsylvania State
           University, University Park, PA 16802, USA
     \and
           Toru\'n Centre for Astronomy, Nicolaus Copernicus University,
           Gagarina 11, 87-100 Toru\'n, Poland
          }

\offprints{O.~L\"ohmer, e-mail: loehmer@mpifr-bonn.mpg.de}
\date{Received / Accepted}

\titlerunning{Cold dust around neutron stars}
\authorrunning{O.~L\"ohmer et al.}

\abstract{We present observations of nine radio pulsars using the
Heinrich-Hertz-Telescope at $\lambda$\,0.87\,mm and the IRAM 30-m
telescope at $\lambda$\,1.2\,mm in search for a cold dust around these
sources. Five of the program pulsars have been observed for the first
time at the mm-wavelengths.  The results are consistent with the
absence of circumpulsar disks that would be massive enough ($\ge
0.01 M_{\odot}$) to support planet formation according to the
scenarios envisioned for solar-type stars, but they do not exclude
lower mass ($\le 10-100 M_{\oplus}$) disks for a wide range of grain
sizes. These conclusions confirm the previously published results and,
together with the current lack of further detections of pulsar
planets, they suggest that planet formation around neutron stars is
not a common phenomenon.

  }

\maketitle

\section{Introduction \label{intro}}

The existence of planets around one of the millisecond pulsars, PSR
B1257+12 (Wolszczan \& Frail 1992\nocite{wf92}; Wolszczan
1994\nocite{wol94}), has raised a possibility that at least some
pulsars may, like normal stars, be accompanied by protoplanetary or
debris disks.  Most theories of planet formation around neutron stars
assume a creation of some sort of a protoplanetary disk out of the
material supplied by the pulsar's binary companion or, possibly, by
the fallback of supernova ejecta (e.g.  Podsiadlowski
1993\nocite{pod93b}; Phinney \& Hansen 1993\nocite{ph93b}).  Of
course, properties and composition of such disks may be significantly
different from those of the disks commonly observed around young
normal stars (e.g.\ Boss 2003\nocite{bos03}). In addition, the initial
conditions for planet formation in an expanding disk must depend on
its ability to protect itself from high-energy photons generated by
early accretion onto the pulsar and from a wind of ultrarelativistic
particles that carries most of the pulsar's spindown energy (Phinney
\& Hansen 1993; Miller \& Hamilton 2001\nocite{mh01}).  Although the
spindown luminosities of pulsars are, in principle, more than
sufficient to heat the dust grains, the efficiency of this process
depends on a degree of coupling of the pulsar wind to dust, which is
difficult to predict in absence of a sufficient observational
evidence.

Since the discovery of the PSR B1257+12 planets, there have been
several attempts to detect the hypothetical circumpulsar disks with
both the space- and the ground-based telescopes.  Upper flux limits
have been derived for a number of sources at wavelengths ranging from
10 $\mu$m to 3 mm (van Buren \& Terebey 1993\nocite{vt93} (IRAF);
Zuckerman 1993\nocite{zuc93} (IRTF); Phillips \& Chandler
1994\nocite{pc94} (JCMT, OVRO); Foster \& Fischer 1996\nocite{ff96}
(IRTF); Greaves \& Holland 2000\nocite{gh00} (JCMT); Koch-Miramond et
al. 2002\nocite{khp+02} (ISO, IRAF/Scanpi); Lazio \& Fischer
2004\nocite{lf04} (ISO)).  As shown by these authors, the existing
infrared and mm-wave flux limits for PSR B1257+12 and other pulsars do
rule out massive, $\ge 0.01 M_{\odot}$ disks similar to those thought
to give rise to planets around normal stars (Boss 2003). However,
these limits do not contradict a possibility that some pulsars may be
accompanied by much less massive disks ranging from a fraction of the
asteroid belt mass to a few hundred $M_{\oplus}$.  In fact, the three
PSR B1257+12 planets of a total mass of $\sim$8\,$M_{\oplus}$ appear
to have formed from a proto\-planetary disk, as suggested by the
observed coplanarity of their orbits (Konacki \& Wolszczan
2003\nocite{kw03}).  Along these lines, Miller \& Hamilton (2001) have
argued that the lower limit to a disk mass necessary to shield the
forming PSR B1257+12 planets against the particle and the photon
emission of the pulsar would be around $\sim$10$^{28}$\,g.
Furthermore, Hansen (2002\nocite{han02b}) has considered PSR B1257+12
planet formation scenarios in which the expanded, planet building part
of the disk would require a mass comparable to that of the existing
planets.

%
\tabcolsep9pt
\begin{table*}[th]
\begin{center}
\caption
     {Continuum flux limits for nine pulsars at $\lambda$\,0.87\,mm and
$\lambda$\,1.2\,mm \label{tab:flux} 
}
\begin{tabular}{lrccrrrc}
\hline\hline\noalign{\smallskip} \multicolumn{1}{c}{PSR} &
\multicolumn{1}{c}{$P$} & {$\dot P$} & \multicolumn{1}{c}{$d$} &
\multicolumn{1}{c}{$\tau$} & \multicolumn{1}{c}{$S_{0.87}$} &
\multicolumn{1}{c}{$S_{1.2}$} & \multicolumn{1}{c}{Remarks} \\ &
\multicolumn{1}{c}{(ms)} & \multicolumn{1}{c}{(10$^{-18}$\,s
s$^{-1}$)} & \multicolumn{1}{c}{(pc)} & \multicolumn{1}{c}{(Myr)} &
\multicolumn{1}{c}{(mJy)} & \multicolumn{1}{c}{(mJy)} & \\

\noalign{\smallskip}\hline\noalign{\smallskip}
\object{J0108$-$1431}\,...... & 807.6 & 77.0 & 130 & 166 & 14.5 & --- & solitary PSR\\
\object{B0950+08}\,.........  & 253.1 & 229.8 & 160 & 17.5 & 14.6 & --- & solitary PSR\\
\object{J1012+5307}\,......  & 5.3 & 0.017 & 520 & 4860 & --- & 2.7 & MSP-WD binary\\
\object{J1024$-$0719}\,......  & 5.2 & 0.019 & 350 & 4410 & --- & 1.8 & solitary MSP \\
B1257+12\,......... & 6.2 & 0.114 & 620 & 863 & 7.6 & 2.0 & MSP, 3 planets \\
\object{B1855+09}\,.........  & 5.4 & 0.018 & 700 & 4760 & 12.5 & 2.0 & MSP-WD binary \\
\object{J2124$-$3358}\,......  & 4.9 & 0.021 & 250 & 3800 & 56.0 & --- & solitary MSP \\
\object{J2307+2225}\,......  & 535.8 & 8.7 & 380 & 976 & 12.0 & --- & solitary PSR \\
\object{J2322+2057}\,......  & 4.8 & 0.010 & 780 & 7850 & 10.2 & --- & solitary MSP \\
\noalign{\smallskip}\hline
\end{tabular}
\vspace{0.2cm}\\ 
\end{center}
{\small Notes.-- (1) Pulsar distances, $d$, were
derived from the Taylor \& Cordes (1993\nocite{tc93}) model of the
Galactic electron density distribution. (2) The characteristic age of
a pulsar is defined as $\tau=P/2\dot P$, where $P$ is the pulsar
period and $\dot P$ is the period derivative. (3) The flux values,
$S_{0.87}$ and $S_{1.2}$, are 2$\sigma$ limits.}
\end{table*}
%
In this paper, we present observations of nine pulsars with the
Heinrich-Hertz-Telescope at $\lambda$\,0.87\,mm and the IRAM 30-m
telescope at $\lambda$\,1.2\,mm. Our measurements were designed to
either detect or to set upper limits to an emission from possible cool
protoplanetary or debris disks around these objects. Five of the
program pulsars have been observed at these wavelengths for the first
time. We describe the instrumental setup and present flux density
measurements of the nine sources in Sect.\ 2. In Sect.\ 3, these
and other results are further discussed in the context of the recent
observational and theoretical developments.

\section{Observations and data analysis \label{obs}}

The sample of pulsars selected for our survey consisted of nine nearby
sources within the distance limit of $<$ 1 kpc. Alongside with the
planets pulsar, PSR B1257+12, we have included three other solitary
millisecond pulsars (MSPs) on the assumption that they would be the
most probable candidates for a detectable dust emission. As binary
MSPs must have undergone long periods of accretion from their
companions and could be left with a residual circumstellar material
suitable for planet formation (Phinney \& Hansen 1993), two MSP-WD
(white dwarf) binaries have been also included. Finally, for
completeness, we have supplemented our list with the three nearby
normal pulsars that would be observable at the telescope sites.
 
The continuum observations at $\lambda$\,0.87\,mm were carried out at
the Heinrich-Hertz-Telescope (Baars et al.\ 1999\nocite{bmm+99}) on
Mt.\ Graham, Arizona, during two observing sessions in January 2001
and December 2002. We used a sensitive 19-channel bolometer array
developed by E.\ Kreysa and collaborators at the Max-Planck-Institut
f\"ur Radioastronomie, Bonn. The centre frequency of the bolometer was
about 345 GHz, with a bandwidth 70 GHz. The beam size on the sky was
$\sim$23$\arcsec$. Using the central channel of the bolometer, we
performed standard on-off measurements with the telescope secondary
chopping in azimuth by $40\arcsec$. In order to calculate the
atmospheric zenith opacity, we made skydip observations every 45
min. The resulting opacities at the observing frequency varied between
0.3 and 0.9 during the two observing sessions. For calibration
purposes we performed mapping and on-off measurements of various
planets (mainly Mars, Saturn and Uranus), resulting in conversion
factors of $7.7-9.5$ mJy per count.

The observations at $\lambda$\,1.2\,mm were carried out in February
and March 2003 at the IRAM 30-m telescope on Pico Veleta (Spain). We
used the 117-channel Max-Planck Millimeter Bolometer (MAMBO-2) array
(Kreysa et al.\ 1999\nocite{kgg+99}), which had a half-power spectral
bandpass from 210 to 290 GHz with an effective frequency of 250
GHz. The beam size on the sky was 10.7 arcsec. The sources were
observed with a single channel using the standard on-off mode with the
telescope secondary chopping in azimuth by $40\arcsec$ at a rate of 2
Hz. The sky opacity at the observing frequency was monitored with
skydips and varied between 0.1 and 0.3 during the seven days of the
observing run. For flux calibration we observed a number of
calibration sources, resulting in conversion factors of $32-37$ counts
per mJy and an estimated absolute flux uncertainty of 15\%.

The data obtained at the two telescopes were converted into a standard
format and analysed using the NIC software package. Correlated noise
was subtracted from the reference channel using the weighted average
signals from the surrounding channels. Anomalous signals were clipped
above the 5$\sigma$ level.

We did not detect dust emission towards any of the pulsars observed. Table 1
lists the 2$\sigma$ flux limits obtained at the two observing
frequencies. At $\lambda$\,0.87\,mm we reached limits of $7-14$ mJy
for most of the sources, which is slightly more than the limits of the
survey by Greaves \& Holland (2000). At $\lambda$\,1.2\,mm 2$\sigma$
flux limits of $1.8-2.7$ mJy were obtained.

%
\begin{figure*}[th]
\centering
\psfig{file=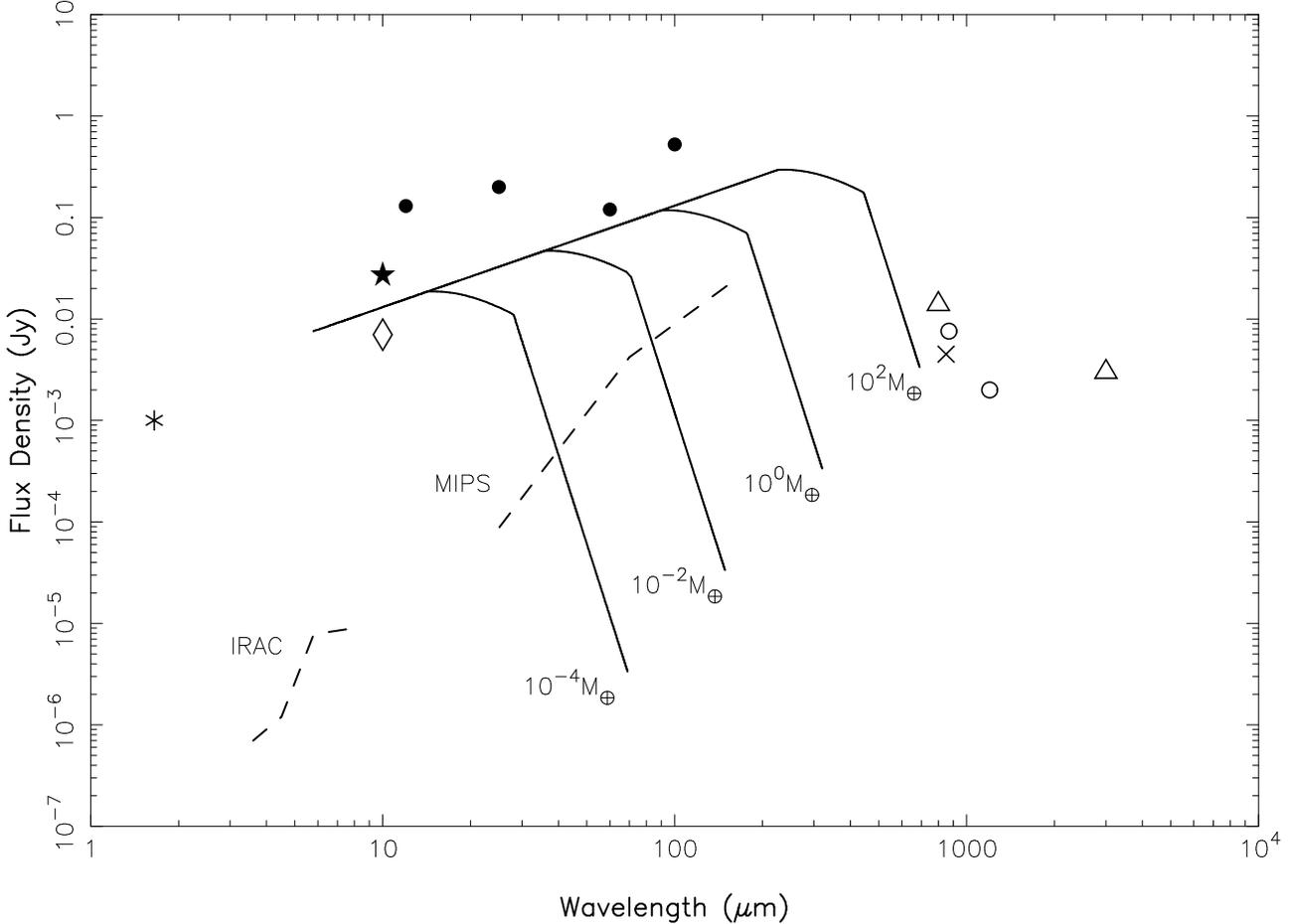,angle=-90,width=10cm}
\caption{ \label{fig:limits} Observational limits and theoretical
models for the emission from dust in the PSR B1257+12 planetary
system.  Symbols mark the flux limits set by observations discussed in
the text ($\circ$) and by published measurements with the following
instruments and surveys: IRTF($\star,\diamond$), IRAS/Scanpi
($\bullet$), JCMT/OVRO ($\triangle, \times$), and 2MASS ($\ast$).  The
solid curves trace the loci of a maximum flux for theoretical disk
emission models calculated for four disk masses as a function of the
grain size ranging from 0.01 to 1000 $\mu$m.  Dashed lines correspond
to the Spitzer Space Telescope 3$\sigma$ sensitivity limits estimated
for the IRAC and the MIPS detectors.  }
\end{figure*}
%

\section{Discussion \label{disc}}

A lack of detections and the derived flux limits at
$\lambda$\,0.87\,mm and $\lambda$\,1.2\,mm for the nine pulsars
discussed in this paper (Table 1) place useful constraints on the
existence of cold, dusty disks around these sources.  To illustrate
the implications of this result for a circumpulsar disk observability
in the framework of a simple model, we use the approach of Foster \&
Fischer (1996) by assuming that a fraction $f$ of the pulsar spindown
luminosity $\dot E$ heats a disk of mass $m_d$ consisting of dust
grains of size $a$ to a temperature $T$, and calculate the expected
infrared flux as a function of disk parameters and pulsar distance
following the formalism developed by these authors.  Because PSR
B1257+12 has planets that have originated in a disk (Konacki \&
Wolszczan 2003), it is the best sampled pulsar in the sub-mm/mm range,
and the published disk mass limits are not dramatically different from
one pulsar to another, we can use this object as a representative
example to make a general comparison of model calculations with the
data.

In Fig.~\ref{fig:limits}, four sets of models for disk masses
$m_d=10^{-4}$, 10$^{-2}$, 10$^0$, and 10$^2 M_{\oplus}$ are
compared with the flux limits for PSR B1257+12 obtained in this work
and the ones published elsewhere.  To ensure compatibility with other
results, we have assumed $k$=0.01 (Foster \& Fischer 1996), 0.01
$\mu$m $\le a\le$ 1000 $\mu$m (Koch-Miramond et al. 2002), the
standard spindown luminosity of the pulsar, $\dot
E=2\times 10^{34}$ erg s$^{-1}$, and the pulsar distance of 620 pc
from the Taylor \& Cordes (1993) model of the Galactic electron
density distribution. In the modelling process, for the assumed disk
mass and grain size ranges, the grain temperature varied over the
range of 8 K $\le T\le$ 500 K, from the largest to the smallest values
of $m_d$ and $a$, respectively. This is practically within the range
between the typical temperature of the cold interstellar dust (10 K)
and the sublimation temperature of silicate grains (1500 K) 
(e.g. Koch-Miramond et al. 2002).

A comparison of the existing observations with a simple dust model
shown in Fig.~\ref{fig:limits} clearly demonstrates the absence of a
disk around PSR B1257+12 with the mass $\ge$~$0.01 M_{\odot}$ that
appears to be necessary to support planet formation around solar-type
stars (Boss 2003). This conclusion is also valid for our data on other
pulsars, and it is generally consistent with the limits on dusty disk
masses obtained from previous observations of a number of pulsars in
the $\lambda$\,10\,$\mu$m -- $\lambda$\,3\,mm range (e.g. Greaves \&
Holland 2000; Koch-Miramond et al. 2002).  Of course, one could alter
this outcome by varying the heating efficiency of the disk by the
pulsar wind, or changing the disk model assumptions. For example, an
upper limit in excess of $0.01 M_{\odot}$ can be obtained for the mass
of a disk around PSR B1257+12 by extrapolating from models of the T
Tauri disks (Phillips \& Chandler 1994). Similarly, a dusty disk model
used by Jura (2003) to explain an infrared excess in the white dwarf
G29-38 can, in principle, be modified to describe the illumination of
a disk by nonthermal radiation from a neutron star (M. Jura, personal
communication). However, these results cannot be conclusively compared
with the ones based on more generic models (e.g. Foster \& Fischer
1996; Koch-Miramond et al. 2002) without a much better understanding
of the physics of hypothetical circumpulsar disks.

As stated earlier and evident from Fig.~\ref{fig:limits}, the data do
not exclude a possibility that dusty disks ranging from a fraction of
the mass of the solar system asteroid belt up to possibly
$\sim$100\,$M_{\oplus}$ may exist around pulsars over a wide range of
temperatures and grain sizes.  Conditions for the existence of such
disks have been recently examined by Miller \& Hamilton (2001), who
specifically predict that a leftover material cannot exist around
pulsars without planets. This is because such pulsars have evidently
not managed to develop protoplanetary disks with a sufficient mass
($m_d\ge 10^{28}$\,g for PSR B1257+12) to protect themselves against
evaporation by the pulsar wind and accretion flux that would prevent
planet formation. On the other hand, pulsars with planets, like PSR
B1257+12, could still have observable debris disks left over from the
final stages of a protoplanetary disk evolution (e.g. Hansen 2002).
In principle, dramatically more sensitive observations of PSR B1257+12
and other pulsars with the Spitzer Space Telescope will have a
potential to either reveal such a circumpulsar debris material or to
place very meaningful constraints on its existence
(Fig.~\ref{fig:limits}).

Our data support a general conclusion drawn from the previous negative
results of searches for dust emission from the vicinity of pulsars
that planet formation around neutron stars occurs infrequently and
that it follows an evolutionary path that must generally be different
from the one envisioned for planets around normal stars.  This
conclusion particularly refers to possible sources of a circumpulsar
matter, as well as to the problem of a disk formation and retention
around a neutron star (Podsiadlowski 1993; Phinney \& Hansen 1993;
Miller \& Hamilton 2001).  It is also consistent with a crude
statistic of $<$ 10\% derived from the fact that PSR B1257+12, one out
of the nine known solitary millisecond pulsars, has planets around it
(Konacki \& Wolszczan 2003). Similarly, it is not surprising that no
massive dusty disks have been found around relatively few normal,
young pulsars observed so far, as the formation and survival of
planets around such objects may be particularly difficult (Thorsett \&
Dewey 1993\nocite{td93}).

In the absence of new detections of planets around millisecond pulsars
(the Jupiter-mass planet in the M4 globular cluster, recently
confirmed by Sigurdsson et al.\ (2003\nocite{srh+03}), was probably
created around a normal star) further searches for dust around these
objects, covering all the physically plausible parameter space at a
possibly high sensitivity level, is a logical way to meaningfully
constrain the models of a creation and evolution of pulsar
protoplanetary and debris disks. Given the existing correlation
between stellar metallicity and the frequency of planets around normal
stars (Santos et al. 2003\nocite{sim+03}; Fischer et al.
2004\nocite{fvm04}), it is likely that the protoplanetary disks around
pulsars, possibly made of a highly evolved stellar material, do not
have to be very massive to support an efficient planet formation (see
also Lazio \& Fischer 2004).  Furthermore, one would also have to
include a distinct possibility that the ``dusty plasma'' environment
of a circumpulsar disk (e.g. Mendis \& Rosenberg 1994\nocite{mero94}),
driven by the relativistic pulsar wind and the magnetic fields carried
with it, cannot be satisfactorily approximated by the standard
protoplanetary or debris disk models based on theories and
observations of the formation regions of solar-type stars (e.g. Boss
2003). In the near future, the Spitzer Space Telescope, with its
factor of $\sim$10$^2 - 10^4$ improvement in sensitivity compared to
the instruments previously used to search for dust emission from
around pulsars (Fig.~\ref{fig:limits}), is an obvious choice for
further exploration of the physics of neutron star planetary systems.

\begin{acknowledgements}

We are very grateful to the staff at the HHT and the 30-m telescope
for their excellent support. We thank M.\ Dumke and M.\ Thierbach for
their help during the observations. 
A.W.'s research was supported by the Alexander von Humboldt
Foundation, the NASA grant NAG5-13620, and by the NSF under Grant
No. PHY99-07949.

\end{acknowledgements}

\bibliographystyle{aa}

\begin{thebibliography}{23}
\expandafter\ifx\csname natexlab\endcsname\relax\def\natexlab#1{#1}\fi

\bibitem[{{Baars} {et~al.}(1999){Baars}, {Martin}, {Mangum}, {McMullin}, \&
  {Peters}}]{bmm+99}
{Baars}, J.~W.~M., {Martin}, R.~N., {Mangum}, J.~G., {McMullin}, J.~P., \&
  {Peters}, W.~L. 1999, \pasp, 111, 627

\bibitem[{{Boss}(2003)}]{bos03}
{Boss}, A.~P. 2003, in ASP Conf. Ser. 294: Scientific Frontiers in Research on
  Extrasolar Planets, 269--276

\bibitem[{Fischer {et~al.}(2004)Fischer, Valenti, \& Marcy}]{fvm04}
Fischer, D., Valenti, J.~A., \& Marcy, G. 2004, IAU Symp.~219: {\it Stars as
  Suns: Activity, Evolution and Planets}, in press

\bibitem[{{Foster} \& {Fischer}(1996)}]{ff96}
{Foster}, R.~S. \& {Fischer}, J. 1996, \apj, 460, 902

\bibitem[{{Greaves} \& {Holland}(2000)}]{gh00}
{Greaves}, J.~S. \& {Holland}, W.~S. 2000, \mnras, 316, L21

\bibitem[{{Hansen}(2002)}]{han02b}
{Hansen}, B.~M.~S. 2002, in ASP Conf. Ser. 263: Stellar Collisions, Mergers and
  their Consequences, ed. M.~M. {Shara}, 221

\bibitem[{{Jura}(2003)}]{jur03}
{Jura}, M. 2003, \apj, 584, L91

\bibitem[{{Koch-Miramond} {et~al.}(2002){Koch-Miramond}, {Haas}, {Pantin},
  {Podsiadlowski}, {Naylor}, \& {Sauvage}}]{khp+02}
{Koch-Miramond}, L., {Haas}, M., {Pantin}, E., {et~al.} 2002, \aap, 387, 233

\bibitem[{{Konacki} \& {Wolszczan}(2003)}]{kw03}
{Konacki}, M. \& {Wolszczan}, A. 2003, \apjl, 591, L147

\bibitem[{{Kreysa} {et~al.}(1999){Kreysa}, {Gem\"und}, {Gromke}, {Possenti},
  {Manchester}, {Camilo}, {McLaughlin}, {Lorimer}, {D'Amico}, {Joshi}, \&
  {Reynolds}}]{kgg+99}
{Kreysa}, E., {Gem\"und}, H.-P., {Gromke}, J., {et~al.} 1999, Infrared Physics
  and Technology, 40, 191

\bibitem[{{Lazio} \& {Fischer}(2004){Lazio}, \& {Fischer}}]{lf04}
{Lazio}, T.~J.~W., \& {Fischer}, J. 2004, astro-ph/0405344

\bibitem[{{Mendis} \& {Rosenberg}(1994)}]{mero94}
{Mendis}, D.~A. \& {Rosenberg}, M. 1994, \araa, 32, 419

\bibitem[{{Miller} \& {Hamilton}(2001)}]{mh01}
{Miller}, M.~C. \& {Hamilton}, D.~P. 2001, \apj, 550, 863

\bibitem[{Phillips \& Chandler(1994)}]{pc94}
Phillips, J.~A. \& Chandler, C.~J. 1994, ApJ, 420, L83

\bibitem[{Phinney \& Hansen(1993)}]{ph93b}
Phinney, E.~S. \& Hansen, B. M.~S. 1993, in ASP Conf. Ser. 36: Planets around
  Pulsars, ed. J.~A. Phillips, S.~E. Thorsett, \& S.~R. Kulkarni, 371--390

\bibitem[{Podsiadlowski(1993)}]{pod93b}
Podsiadlowski, P. 1993, in ASP Conf. Ser. 36: Planets around Pulsars, ed. J.~A.
  Phillips, S.~E. Thorsett, \& S.~R. Kulkarni, 149--165

\bibitem[{{Santos} {et~al.}(2003){Santos}, {Israelian}, {Mayor}, {Rebolo}, \&
  {Udry}}]{sim+03}
{Santos}, N.~C., {Israelian}, G., {Mayor}, M., {Rebolo}, R., \& {Udry}, S.
  2003, \aap, 398, 363

\bibitem[{{Sigurdsson} {et~al.}(2003){Sigurdsson}, {Richer}, {Hansen},
  {Stairs}, \& {Thorsett}}]{srh+03}
{Sigurdsson}, S., {Richer}, H.~B., {Hansen}, B.~M., {Stairs}, I.~H., \&
  {Thorsett}, S.~E. 2003, Science, 301, 193

\bibitem[{Taylor \& Cordes(1993)}]{tc93}
Taylor, J.~H. \& Cordes, J.~M. 1993, ApJ, 411, 674

\bibitem[{Thorsett \& Dewey(1993)}]{td93}
Thorsett, S.~E. \& Dewey, R.~J. 1993, ApJ, 419, L65

\bibitem[{{van Buren} \& {Terebey}(1993)}]{vt93}
{van Buren}, D. \& {Terebey}, S. 1993, in ASP Conf. Ser. 36: Planets around
  Pulsars, ed. J.~A. Phillips, S.~E. Thorsett, \& S.~R. Kulkarni, 327--333

\bibitem[{Wolszczan(1994)}]{wol94}
Wolszczan, A. 1994, Science, 264, 538

\bibitem[{Wolszczan \& Frail(1992)}]{wf92}
Wolszczan, A. \& Frail, D.~A. 1992, Nature, 355, 145

\bibitem[{{Zuckerman}(1993)}]{zuc93}
{Zuckerman}, B. 1993, in ASP Conf. Ser. 36: Planets around Pulsars, ed. J.~A.
  Phillips, S.~E. Thorsett, \& S.~R. Kulkarni, 303--315

\end{thebibliography}

\end{document}